\documentclass[twocolumn,showkeys,showpacs,aps,nofootinbib]{revtex4}
\usepackage{graphicx}
\usepackage{dcolumn}
\usepackage{bm}
\usepackage{amsmath}

\newcommand{\reff}[1]{(\ref{#1})}
\newcommand{\eref}[1]{Eq.(\ref{#1})}
\newcommand{\erefs}[1]{Eqs.(\ref{#1})}

\newcommand{\qq}{\qquad}
\newcommand{\vecc}[1]{\boldsymbol{#1}}
\newcommand{\Bm}{\bar{B}}
\newcommand{\psiM}{\bar{\psi}}
\newcommand{\etaV}{\eta_v}
\newcommand{\psz}{\psi_0}
\newcommand{\psu}{\psi_1}
\newcommand{\pzb}{\bar{\psi}_0}
\newcommand{\pub}{\bar{\psi}_1}
\newcommand{\putt}{\tilde{\psi}_1}
\begin{document}

\title{Inconsistency in the Standard Model for Stellar Thin Accretion Disks}

\author{Giovanni Montani$^{a,b}$, Nakia Carlevaro$^b$}
\affiliation{\vspace{3mm}
$^a$ ENEA - C.R. Frascati (Rome), UTFUS-MAG.\vspace{1mm}\\
$^b$ Physics Department, ``Sapienza'' University of Rome,\\
\emph{\footnotesize c/o VEF, ``Sapienza'' Universit\`a di Roma}\\
\emph{\footnotesize P.le Aldo Moro 5, 00185 Roma (Italy)}.}

\date{\today}
\begin{abstract}
We analyze the configuration of a thin rotating accretion disk, which is embedded in a magnetic field inducing a backreaction in the gravitating plasma. The aim of this study is to determine the conditions under which the gaseous accretion model of Shakura can be reconciled with the magneto-hydrodynamical picture requested to trigger the underlying turbulent behavior. We focus our attention to the generalized Ohm equation in order to understand if the plasma backreaction is able to provide the proper toroidal current, allowing a non-zero infalling velocity. In the limit of linear plasma backreaction, this analysis shows how the Shakura profile of accretion turns out to be inconsistent. In particular, comparing the azimuthal and the generalized Ohm equilibrium equations, we argue that it is not possible to maintain a constant rate of accretion. A non-stationary scenario for the disk configuration is then outlined and it results into a transient process which is however associated to a vanishing accretion rate.
\end{abstract}

\maketitle

\section{Introduction}
One of the fundamental processes, governing the behavior of a wide class of astrophysical sources, is the accretion of matter onto a compact object \cite{BKL01}. The material surrounding massive accretors is, in general, in the state of a completely ionized plasma, as a consequence of both the high temperature and low density conditions. Nonetheless, the first proposal to describe the accretion process, due to Shakura in the early seventies \cite{S73, SS73}, was based on a hydrodynamical approach for the infalling of a gas-like matter. This simplification was necessary to shed light on the mechanism underlying the angular momentum transport. This aim was successfully reached identifying the shear viscosity, associated to the differential rotation of the disk, as the driving phenomenon.

The Shakura prescription is still widely adopted as a basic estimation for the accretion rate of astrophysical system versus viscosity. Indeed, despite recent multidimensional simulations \cite{BY10, P11, SS11, McK12} do not directly rely on the Shakura idea, nonetheless this model has still a timely profile both when theoretical \cite{TM11} and numerical \cite{B08, MFZ10} analyses are addressed. Since stellar accretion disks are mainly characterized by thin equilibrium configurations, the use of the one-dimensional model appears a well-grounded approximation and the comparison with detailed simulations is certainly viable, especially in view of the resulting values for the Prandtl number. The implementation of the Shakura prescription relies on its capability to offer a reliable paradigm for observations. Nonetheless, the microscopic features of the accreting plasma are typically compatible with a quasi-ideal system, for which the shear viscosity coefficient takes values much smaller than those ones requested to account for the observations. Shakura, being aware of this discrepancy, proposed that the dissipative effects arise from a quasi-ideal turbulent behavior, restated as an effective viscous laminar flow.

The solution to the puzzling picture concerning the linear stability of an axisymmetric rotating disk (incompatible with the onset of turbulent regimes) was offered by the astrophysical implementation that Chandrasekar gave \cite{Ch60} to the so-called magneto-rotational instability (MRI), just derived by Velikov \cite{V59} but in a slightly different context. The merit of having understood how the MRI could be the proper mechanism to generate a dissipative dynamical regime, is due to Balbus and Hawley \cite{BH91} (the analysis of alternative mechanisms responsible for unstable plasma modes in the disk can be found in \cite{CC01, Co08}). Indeed, they clarified the role that an arbitrarily small magnetic field can play in developing an efficient turbulent behavior when coupled to a differentially rotating plasma. The crossmatching of the Shakura idea of angular momentum transport with the MRI (here named the Standard Model for accretion), triggering of a turbulent regime, appears as a solid theoretical framework consistent with a significant set of observations \cite{BKL01, BH98}.

The present analysis is focused on a relic puzzle contained in such a picture (the so-called anomalous resistivity of the disk plasma \cite{MB11}) and on the suggestion that a revision of the basic hydrodynamical Shakura idea is mandatory. Indeed, including a magnetic field in the equilibrium configuration is a natural perspective when dealing with a real astrophysical source, often endowed with a significant magnetosphere, but it implies that the plasma (or simply the magneto-active fluid) backreaction must be accounted too. In particular, due to the axial symmetry, the generalized Ohm law (GOL) requires that a relevant resistivity coefficient is postulated in order to deal with a non-zero radial infall velocity. In fact, the current density arising in the plasma in view of its backreaction to the central object field (which has no associated intrinsic current since it is a vacuum field and it is approximated here by a dipole profile) can not be, in general, sufficiently strong to guarantee the balance accounting for the very small microscopic value of the plasma resistivity. In this paper, rather than investigate the mechanisms generating the anomalous values of the resistivity, we show that this puzzle enclose an inconsistency of the Shakura idea of accretion, when the full magneto-hydrodynamical (MHD) scheme is addressed. As natural development of our configuration scheme, we analyze a non-stationary profile of the disk which however turns out to provide a vanishing accretion rate on the compact object.

The new feature of the one-dimensional model we present here (concerning also higher-dimensional numerical simulations) is the proof that the Shakura idea for the angular momentum transport holds only in the presence of anomalous resistivity. This way, such a quantity becomes a theoretical prescription more than a phenomenological request. We reach this issue by a detailed balance of the magneto-static plasma backreaction and, apart from non-trivial technicalities, it is expected to hold in more than one-dimension too. The difficulty we outline here in reconciling the plasma magnetic backreaction with the Shakura prescription is particularly interesting in view of recent experimental \cite{PL11} and numerical \cite{A12} issues which suggest the non-linear instability of neutral fluid in Keplerian rotation (see also \cite{B11}), as well as in view of the theoretical analysis \cite{CTM12} which states (on a kinetic level) the stability of a magnetized plasma having a certain temperature anisotropy (against the MRI paradigm).

The necessity for a reformulation of the accretion process was proposed in \cite{Co94, Co07} (see also \cite{BM09}), where a parallelism between the axial symmetric configuration of Tokamak fusion reactors and accretion disks was proposed. As underlying picture of the new point of view, in \cite{Co05, CR06} (see also the global extension provided in \cite{MB11}), it has been determined the existence of a crystal-like (\emph{i.e.}, a radially oscillating profile) configuration for the magnetic field of a thin purely rotating disk. This scenario consists of a radial periodic structure of the magnetic flux surfaces, that, in the limit of a strong backreaction of the plasma, leads to a fragmentation of the profile into a ring series (for an analysis of jet formation in this scheme, see also \cite{MC10}). The main merit of this new proposal is to outline how the Lorentz force, associated to the plasma backreaction, can play a relevant role in fixing the configuration, in view of the small-scale structure it acquires in the disk. More specifically, the Lorentz force is the proper contribution able to balance (already in the case of a linear backreaction) the higher order correction to the Keplerian centripetal force, in place of the gravitational effects. Indeed, the balance between the centripetal and gravitational forces takes place when identifying the angular velocity of the disk with its Keplerian profile at all orders of approximation in the mass density. Therefore, the additional term induced in the centripetal force by the backreaction must be included in the magneto-static equilibrium. Since also the Lorentz force is due to the plasma backreaction, it results natural to require that both the centripetal correction and the Lorentz term stand at the same equilibrium scale, as in \cite{Co05}. It is just this feature that makes our equilibrium configuration different from previous solutions of the thin disk problem (see \cite{B03, BKL01, SITS04, GO12, MFZ10, PCP06}). This is the line of thinking at the base of the present treatment for accretion process. 

The paper is organized as follows: in Section \ref{paradigm}, we discuss the fundamental framework in which our analysis is performed. We introduce the fundamental quantities and the concept of vertical average necessary to reduce the dynamics to a one-dimensional problem. In Section \ref{funeq}, we construct the fundamental equations governing the equilibrium configuration. In particular, we recognize in the radial equation a separation between the hydrodynamical and the magnetic backreaction scales. Section \ref{inshak} is devoted to outline the inconsistency of the Shakura model in the presence of a magnetic field by a detailed analysis of the azimuthal momentum conservation equation. We also show how an analytical solution, radially oscillating and damped in time, can be obtained but it corresponds to a net zero accretion rate of the disk. We then conclude this section providing an estimate showing how the only viable approach to the Shakura prescription implies an anomalous resistivity coefficient. In Section \ref{concl}, we present a physical discussion concerning the nature of the magnetic backreaction at the ground of our approach. An astrophysical characterization of our model is offered in order to focus the nature of the obtained inconsistency.

\section{The underlying paradigm}\label{paradigm}

We now construct a non-stationary MHD model for the equilibrium of a thin accretion disk, surrounding a compact astrophysical object. In axial symmetry, the magnetic field $\vecc{B}$ reads as
\begin{align}
\vecc{B}=-\hat{\vecc{e}}_r\,\partial_z\psi/r+\hat{\vecc{e}}_z\,\partial_r\psi/r\;,
\end{align}
in view of its divergenceless nature ($(r,\phi,z)$ being cylindrical coordinates). Here, we set to zero the toroidal component $B_{\phi}$, while $\psi = \psi(t,r,z^2)$ denotes the magnetic flux surface. Let us now split this function as
\begin{align}
\psi(t,r,z^{2})\simeq\psz(r,z^2)+\psu(r,t,z^2)\;,
\end{align}
denoting by $\psz$ the flux surface associated to the background magnetic field (proper of the central object), and by $\psu$ the contribution due to the backreaction of the disk plasma. In what follows, the flux surface of the background field is taken in a reliable dipole form and we assume a linear response of the plasma $\psu\ll\psz$, for which the induced magnetic field remains small with respect to the background one, \emph{i.e.}, $\partial_r\psz\gg\partial_r\psu$. The density current can be split in the form $\vecc{J}=\vecc{J}_0+\vecc{J}_1$, where $\vecc{J}_0=0$, as requested by the vacuum nature of the corresponding field. Furthermore, the $z$-dependence of $\psu$ can be neglected when calculating the backreaction component $\vecc{J}_1$ of the current. In fact, considering the regime where the $\beta$-parameter of the plasma is much greater than unity, namely $\beta \simeq v_s^2/v_A^2\gg 1$ ($v_s$ and $v_A$ being the plasma sound and Alfv\`en velocity, respectively), the typical vertical scale of the configuration $H\simeq v_s/\Omega$ (where $H=H(r)$ is the half-depth of the disk) is much greater than the corresponding radial one $\lambda \simeq v_A/\Omega$ \cite{Co05} (the consistency of this scheme can be verified \emph{a posteriori} from \erefs{system}). This hierarchy of the typical configuration scales is naturally expected to be maintained in the linear backreaction regime and it allows to neglect the contribution that the $z$-derivatives of $\psu$ would provide to the toroidal current density.

In agreement to the Standard Model for accretion \cite{BKL01}, let us now consider an essentially one-dimensional axisymmetric scheme, in which the $z$-dependence is averaged out using 
\begin{align}
\bar{f}(r,t)=\frac{1}{2H}\;\int^{+H}_{-H}\!\!\!\!
\!\!\!\!f(r,t,z^{2})\,dz\;,
\end{align}
where $f$ denotes a generic coordinate function and we disregarded odd $z$-quantities because their average vanish by definition. This simplification is due to the thinness of the disk and the $z$-derivatives of all the even quantities with respect to the equatorial plane can be neglected together with the vertical component $v_z$ of the fluid velocity (vanishing on the equatorial plane). Accordingly, the mass density $\rho$ can be replaced by the superficial density defined by $\Sigma(t,r)\equiv 2H\bar\rho$, where the thinness condition $H/r\ll1$ holds. While, averaging out the configuration equations, we replace the radial velocity $v_r$ by a mean value $u_r$ weighted over the density and the same for its derivatives (or products of these quantities), \emph{i.e.},
\begin{align}
u_r \equiv \frac{1}{\Sigma}\int_{-H}^{+H}\!\!\!\!
\!\!\!\!\rho v_r dz\;,\qquad
\partial_{t,r}\,u_r\equiv\frac{1}{\Sigma}\int_{-H}^{+H}\!\!\!\!
\!\!\!\!\rho\;\partial_{t,r}\,v_r dz\;.
\end{align}
respectively. Another important property characterizing the disk configuration is its angular velocity $\Omega$, taken equal everywhere in the disk to its equatorial value. 

Introducing the disk accretion rate $\dot{M} = -2\pi r\Sigma u_r$ \cite{BKL01}, the continuity equation writes
\begin{equation}
\partial_t\Sigma-\partial_r\dot{M}/2\pi r=0\;.
\label{avcq}
\end{equation}
We note that, while $\partial_z v_z$ does not vanishes on the equatorial plane, the total $z$-derivative in this equation provides a very small edge term after integration.

After averaging over the vertical profile, the only surviving component of the magnetic field $\vecc{B}$ is 
\begin{align}
\vecc{\Bm} =\hat{\vecc{e}}_z\;\partial_r\psiM/r\;,\qq
\psiM(t,r)=\pzb+\pub\;,
\label{bfieldm}
\end{align}
where $\pzb=\mu/r$ (with $\mu=const.$), coherently with the dipole expression. Averaging now the current density $\vecc{J}=\vecc{J}_1$, we stress that the key point of the proposed scheme consists in the large value that the resulting backreaction current component, \emph{i.e.},
\begin{equation}
\vecc{\bar{J}}_1=
-\frac{c}{4\pi}\;\partial_r\Big(\frac{1}{r}\,\partial_r\pub\Big)\;\hat{\vecc{e}}_{\phi}
\simeq -\frac{c}{4\pi r}\;\partial^2_r\pub\;\hat{\vecc{e}}_{\phi}\;,
\label{toca}
\end{equation}
takes in view of the small scale $\lambda$ of the plasma backreaction (in fact, in the last passage above, we made use of the inequality $\lambda\ll r$ so neglecting the derivative of the $1/r$ factor in front of $\partial_r\pub$).

\section{Fundamental equations}\label{funeq}
Let us now consider the system provided by the Faraday equation and GOL, \emph{i.e.},
\begin{align}
\label{flfb}
\partial_t\vecc{\bar{B}}+c\nabla\times\vecc{\bar{E}}=0\;,\\
c\sigma\,\vecc{\bar{E}}+\sigma\vecc{u}\times\vecc{\bar{B}}-c\vecc{\bar{J}}=0\;,
\label{flfb2}
\end{align}
respectively \cite{BH98}, where $\vecc{\bar{E}}$ denotes the electric field and $\sigma$ the electric conductivity.  
The system above can be easily combined using the expression \reff{bfieldm} for the magnetic field. In particular, the $z$-component of \eref{flfb2} provides $\bar{E}_z=0$, while the radial one writes
\begin{subequations}
\begin{align}
\label{fb2c}
\bar{E}_r\equiv-\partial_r\Phi=-(\Omega/c)\,\partial_r\bar{\psi}\;,
\end{align}
where $\Phi$ denotes the electrostatic potential. In this scheme, the $z$-component of \eref{flfb} yields to the relation $\bar{E}_{\phi}=-\partial_t\pub/cr$, while the two other ones identically vanish. Finally, the toroidal component of \eref{flfb2} writes
\begin{align}
\partial_t\pub+u_r\,\partial_r\pzb=
-cr\,\bar{J}_{1(\phi)}/\sigma\;,
\label{fb2c2}
\end{align}
\end{subequations}
We stress that \eref{fb2c2} is equivalent, apart from a global radial derivative, to the $z$-component (the only non-vanishing one) of the magnetic-field induction equation obtained substituting \eref{flfb2} into \eref{flfb}. 

We observe that \eref{fb2c} is compatible with the assumption $\Omega=\Omega(\bar{\psi}(t,r))$. Since in the absence of backreaction the disk rotation is Keplerian, \emph{i.e.}, 
\begin{align}
\Omega_0=\Omega_K\equiv\sqrt{GM_S/r^3}=\sqrt{GM_S\pzb^3/\mu^3}\;,
\end{align}
($M_S$ being the mass of the central object), the simplest expression satisfying the corotation condition above and able to recover the Keplerian nature at the lowest order can be postulated as
\begin{equation}
\Omega(\bar{\psi})=\sqrt{GM_S\bar{\psi}^3/\mu ^3}
\simeq\Omega_K\Big(1+3\pub/2\pzb\Big)\;,
\label{anve}
\end{equation}
where it is essential that the amplitude of the perturbed magnetic surfaces remains always smaller than the background value $\pzb=\mu/r^3$. In the same approximation scheme, setting $\eta_B=c^2/4\pi\sigma$ and using the relation between $u_r$ and $\dot{M}$, \eref{fb2c2} rewrites
\begin{equation}
\Sigma\partial_t \pub +
\dot{M}\mu/2\pi r^3 =
\Sigma \eta_B\partial^2_r\pub\;.
\label{fbff}
\end{equation}

The radial equilibrium equation \cite{BKL01} can be split into three components, corresponding to different orders of the equilibrium configuration. The first two naturally enter the Shakura model for a non-magnetized fluid, while the third one is due to the plasma backreaction and it balances the corrections of the centripetal force by the Lorentz one. They write
\begin{subequations}
\begin{align}
\label{raeq}
\Omega_0&=\Omega_K\;,\\
\label{raeq2}
\Sigma(\partial_t u_r+u_r\partial_r u_r)&=-\partial_r\Pi\;,\\
\label{raeq3}
\partial^2_r\pub &=-(3GM_s/\mu^2H)\Sigma r^3 \pub\;,
\end{align}
respectively, where $\Pi=2H\bar{p}$ ($p$ denoting the thermostatic pressure). The possibility to restate the radial equilibrium in these three separated force balance equations is a fundamental step of our approach. While the Keplerian nature of the background configuration is a well posed assumption for a thin disk, see \cite{Og97}, the splitting between the hydrodynamical term and the magnetic backreaction can take place only if the plasma $\beta$-parameter satisfy the condition $\beta\ll L/\lambda$ (hence $\lambda\ll L$) being $L$ the variation scale of the pressure. Furthermore, using $\dot{M}$ and \eref{avcq}, \eref{raeq2} rewrites
\begin{equation}
\partial _t\dot{M} +
\partial _r(\dot{M}u_r) = 2\pi r\,\partial_r\Pi\;.
\label{srer}
\end{equation}
\end{subequations}

The vertical equilibrium equation \cite{BKL01} fixes, via the gravostatic equilibrium, the $z$-dependence of the mass density $\rho$. The features of the vertical equilibrium, sensitive to the equation of state, are not needed for the addressed task, apart from fixing $H(r)\simeq v_s/\Omega_K \ll r$.

Finally, using the expression \reff{anve} for $\Omega$, the azimuthal equilibrium equation \cite{BKL01} writes, in the presence of a magnetic field,
\begin{align}
\label{azeqqqq}
\Sigma \partial _t\pub &-\dot{M}\partial _r(r^2\Omega _K)/2\pi=\nonumber\\
&\qquad=\partial_r(2H\etaV r^3\partial_r\Omega_K)+2H\etaV\partial^2_r\pub\;,
\end{align}
where $\etaV$ denotes the shear viscosity coefficient associated to the disk differential rotation (we have neglected the gradient of $\pub$ with respect to its second $r$-derivatives).

\section{Inconsistency of the magnetized Shakura model}\label{inshak}

We remind that the Shakura model \cite{S73} is based on the idea that the accretion rate of the disk is driven by the background shear viscosity. Requiring that the accretion rate is provided by the usual Shakura expression, \emph{i.e.}, $\dot{M}_{Sh}=6\pi H\etaV$ \cite{BKL01}, indeed we are implementing the idea that the magnetic field of the central object has the only role to allow for the MRI. This way, the Shakura model is preserved because the turbulence arising from the MRI induces the effective values of the dissipation coefficients on which the phenomenological predictability of the accretion rate $\dot{M}_{Sh}$ is based. The source of inconsistency can be easily identified in the new equation arising in the MHD scenario, \emph{i.e.},  GOL, especially its azimuthal component. It is just this new equation which can not be reconciled  with the azimuthal component of the momentum conservation \reff{azeqqqq}, once the Shakura accretion rate is imposed.

In fact, following this prescription in order to recover the standard behavior, we assume the decomposition of \eref{azeqqqq} in two equations, one for the background regime and the other one for the backreaction response, \emph{i.e.},
\begin{subequations}
\begin{align}
\label{azeq}
\dot{M}\partial _r(r^2\Omega _K)/2\pi
+\partial _r \left( 2H\etaV r^3
\partial _r\Omega _K\right)=0\;,\\
\label{azeqB}
\Sigma\,\partial _t\pub-2H\etaV\,\partial^2_r\pub=0\;,
\end{align}
\end{subequations}
respectively. \eref{azeq} matches the standard Shakura behavior for the angular momentum transport. In fact, if we focus on the stationary configuration by setting to zero all the partial time derivatives, it reduces to the ordinary Shakura relation $\dot{M}_{Sh}=6\pi H\etaV$. Moreover, \eref{avcq} gives the fundamental constraint $\dot{M}=const.$ 

In the stationary regime, the inconsistency of the Standard Model for accretion disks comes out immediately: \eref{azeqB} requires $\partial^2_r\pub=0$ implying, by virtue of GOL \reff{fbff}, $\dot{M}=0$, \emph{i.e.}, a vanishing disk accretion rate. By other words, in the presence of a magnetic field, a backreaction current of the plasma is mandatory in order to balance the accretion rate in \eref{fbff}, but such a current must vanish (or, more realistically, it must be negligible) to safe the accretion-rate constant profile of the Shakura prescription. This type of inconsistency can not be removed even retaining the $z$-dependence of $\pub$.

We now focus on the non-stationary sector inferring that the background term proportional to viscosity can be neglected, while the accretion process should be driven by the toroidal density current. Thus, \eref{azeqqqq} rewrites
\begin{equation}
\Sigma \partial _t\pub 
-\dot{M}\mu/6\pi r^3
= 2H\etaV\partial^2_r\pub\;,
\label{azeq2}
\end{equation}
and combining \erefs{fbff}, \reff{raeq3} and \reff{azeq2}, we get
\begin{align}
2\dot{M}\mu/3\pi r^3=(2H\etaV-\Sigma\eta_B)
3GM_S\Sigma r^3\;\pub/H\mu ^2\;,
\label{fbff2}\\
\partial_t\pub=
-(2H\etaV + \Sigma\eta_B/3)9GM_Sr^3\;\pub/4H\mu^2\;.
\label{fbff3}
\end{align}

The dynamical system involves four unknowns $\Sigma$, $\dot{M}$, $\pub$ and $\Pi$ obeying \erefs{avcq}, (\ref{srer}), (\ref{fbff2}) and (\ref{fbff3}) (completed retaining \eref{raeq3} too). In the case of linear backreaction, the superficial density must admit the perturbed decomposition $\Sigma=\Sigma_0(r)+\Sigma_1(t,r)$ ($\Sigma_1\ll\Sigma_0$) where $\Sigma _0$ is associated to the background profile, while $\Sigma_1$ is the small time-dependent correction due to the plasma backreaction. In order to get a physical insight on the dynamical content of the accretion paradigm traced above, we can make some reasonable simplifying assumptions setting \cite{MB11}
\begin{subequations}
\begin{align}
\Sigma&=\Sigma_0=2Hm/r^3\;,\\
\eta_B&=const.\;,\\
2H\etaV&=\Sigma\eta_B/n_P\;,
\end{align}
\end{subequations}
where $m=const$, and $n_P=const$ denotes the Prandtl number. The behavior of $\pub$ is now described by
\begin{subequations}\label{system}
\begin{align}
\label{fexi}
\partial^2_r\pub+k^2\pub&=0\;,\\
\partial_t\pub+(\tfrac{1}{4}+\tfrac{3}{4n_P})\eta_B k^2\,\pub&=0\;,\\
\label{knostra}
k^2&\equiv 6GM_Sm/\mu^2\;.
\end{align}
\end{subequations}

The solution reads as
\begin{equation}
\pub=\putt \sin(kr)\;\exp
\big[-(\tfrac{1}{4}+\tfrac{3}{4n_P})\,\eta_B k^2\;t\big]\;,
\label{sofi}
\end{equation}
where $\putt$ is an integration constant and substituting this expression into \eref{fbff2}, we finally get
\begin{equation}
\dot{M} = (1/n_P-1)\,(3\pi/\mu)\,m H\eta_Bk^2\;\pub\;.
\label{sofi2}
\end{equation}
Although the resulting $\dot{M}$ is comparable in amplitude to $\dot{M}_{Sh}$, we stress that the radial infalling velocity is described by the behavior $u_r\propto-r^2\pub$. Since $kr\gg1$ by hypothesis ($k\sim1/\lambda$), the radially oscillating term of $\pub$ varies much rapidly than $r^2$ and, therefore, the last can be approximated as constant when estimating the accretion profile. As a consequence, the obtained non-stationary model provides a radial infalling velocity having a vary fast radial sinusoidal behavior damped in time by the resistive effects. Clearly, such a profile is not associated with a real infalling of material on the central object, but it is essentially a local accretion process which accumulates material both from outside and inside, around a fixed radial coordinate. The same scenario has been also outlined in \cite{MB11a, BMP11} within a two-dimensional local configuration.

Our work claims the impossibility to reconcile the Shakura Model with a fully consistent MHD scheme in which the plasma backreaction can not be neglected. The open question is then if the assumption $kr\gg1$ (\emph{i.e.}, $\lambda\ll r$) at the base of our analysis is really obliged or it can be somehow avoided. We elucidate this point by a chain of estimates based on the fundamental equilibrium equations. In the stationary regime, if we adopt the Shakura relation for the viscosity $\eta_v=2\alpha\Sigma_0v_s/3$ (where $\alpha\le1$ is a parameter of the model), $\dot{M}_{Sh}$ provides the radial infalling velocity $u_r\sim\alpha v_s H/r$. At the same time, \eref{flfb2} gives $u_r\sim B_1 c^{2}/B_0\sigma\lambda$ where $B_0$ and $B_1$ are the background and backreaction magnetic field amplitudes, respectively, and we recall that $\lambda$ denotes the magnetic backreaction scale. Using now the definitions of $\sigma$ and $\eta_B$, and comparing the obtained expressions of the radial velocity, one can easily get 
\begin{align}\label{stima}
\frac{B_1}{B_0}\;\frac{r}{\lambda}\sim\frac{1}{n_p}\;.
\end{align}
Thus, in the linear regime ($B_1\ll B_0$), if we require that the backreaction currents induced in the plasma are not relevant, \emph{i.e.}, $\lambda\sim r$, in order to avoid a vanishing accretion rate in the Standard Model, one must address huge values of the resistivity coefficient since $n_p\gtrsim1$.

This work demonstrates that the Shakura model is inconsistent in the presence of a small-scale magnetic backreaction able to account for an enough strong toroidal current. Indeed, only such a current allows a balance of the electron force in the presence of quasi-ideal (kinetic) values of the resistivity coefficient. If we try to avoid this inconsistency by considering large-scale backreaction, then we have to deal with a very small toroidal current which can be neglected everywhere in setting the equilibrium configuration, but in the toroidal GOL. In fact, in such an equation the term $\eta_B\bar{J}_{1(\phi)}$ must be necessarily retained to balance the Lorentz force acting on the electrons. As a consequence, if the backreaction is of very large-scale and the corresponding toroidal current are tiny, the only way to preserve the consistence of the stationary GOL is to consider a huge value of the coefficient $\eta_B$, \emph{i.e.}, to deal with the so-called anomalous resistivity. This statement is well-elucidate by the estimate above.

Putting together the present analytical treatment, which leads to the solution \reff{sofi}, with the qualitative features outlined by the estimate \reff{stima}, we are able to get light on the origin of the anomalous resistivity puzzle: when the numerical simulations are compared with the observations, a strong value of the resistivity coefficient is needed because very small-scale magnetic backreaction is forbidden by the compatibility of the equilibrium configuration.

\section{Physical hints and final remarks}\label{concl}

It is now worth clarifying the real nature of the plasma backreaction we are referring to and which is the physical meaning of the solution \reff{sofi}. Indeed, the presence of the external magnetic field is responsible for a series of correlated non-linear processes whose net effect is the emergence of the effective visco-resistive properties \cite{B03}. Such phenomena are called plasma backreaction too and it is crucial to stress how it intrinsically differs from the disk reaction we are accounting for.

The fundamental feature to be focused in the plasma backreaction is the interplay which takes place in the disk between the MRI-triggered turbulence (responsible for the magnetic field enhancement) and the turbulent reconnecting transport (transferring outward such a field) \cite{deG12, S-L12a, S-L12b, S-L11}. By other words, the dynamo effect produces a growth of the magnetic field amplitude, which would stop the MRI instability, but the latter is continuously restarted by the decrement of the magnetic field due to the turbulent transport of its flux in the outer regions of the disk, and in the disk corona. We can estimate the typical time scale of these processes in terms of the characteristic growth rate for the MRI instability which, in the best case, is $\tau_{MRI}\sim1/\Omega_K$ \cite{BH91}. Finally, it is worth noting that the essential background for such instability is the purely rotating Keplerian disk, lacking of any accretion feature.

The backreaction scheme we present here, and which is responsible for the inconsistency of the Shakura prescription in the MHD sector, has a very different origin. In fact, it takes place on much shorter time scale than the MRI triggered turbulence and it arises nearby a background configuration which, in addition to the stripped rotating disk possesses, possesses an hydrodynamical scale accounting, in principle, for an accretion process. By other words, the net effect of the complicated interplay between the turbulent dynamo and reconnecting transport is just the emergence of the non-zero accretion of the disk, that we assume at the ground of our analysis. In this sense, the above mentioned processes are here summarized in the non-zero viscosity and resistivity coefficients.

The characteristic time scale of the backreaction we analyzed, as it results from equation \reff{sofi}, is of the order $\tau_{BR}\sim n_P/\eta_B k^2$, where we excluded the limit $n_P\gg1$. Since we are implementing our backreaction on a standard (effective) visco-resistive scenario, we can adopt the value of $\eta_B$ when the Shakura prescription for the viscosity coefficient is considered, \emph{i.e.}, $\eta_B\sim n_P \alpha v_s H \sim n_P \alpha H^2\Omega_K$. Hence, we get the fundamental relation
\begin{equation}
\tau_{BR}\sim\frac{1}{\alpha \omega _K(kH)^2} \sim 
\frac{1}{\alpha (kH)^2}\;\tau_{MRI}\;.
\label{stimt}
\end{equation}

Although the parameter $\alpha$ can take small values and its range of variation is from $\mathcal{O}(10^{-3})$ to $\mathcal{O}(1)$, the quantity $kH$ can take huge values in the disk. An estimation for the wave-number \reff{knostra} has been provided in \cite{MB11} and it gives a minimum value of about $k_{min}\sim10Km^{-1}$. Thus, if we take for the disk half-depth the typical value $H\sim 10^3Km$, even in the best case $\alpha \sim 10^{-3}$, we get the relation $\tau_{BR}\sim 10^{-5}\tau_{MRI}$. Such a striking difference in the characteristic time scale of the MRI triggered turbulence and the backreaction here analyzed outlines the correspondingly different nature between these dynamical processes. The very extremely small period of time, over which the backreaction driven by $\bar{\psi}_1$ is damped, confirms the inconsistency of the Shakura model in the presence of a magnetic field. In fact, there is no way to deal with a longstanding current in the disk, able to account for the proper balance of GOL. This current is induced by the plasma disk backreaction and it is almost instantaneously damped by the dissipative effects (the net result of the turbulent processes on which the Shakura paradigm relies on). As a consequence of this picture, the solution \reff{sofi} more than representing a physical regime of the disk dynamics (to be analyze in view of its possible implications), it results a rapid damping profile of any radial oscillating behavior arising in the disk plasma by means of the balance between the induced centripetal and Lorentz force, respectively. In this sense, the crystalline morphology of the disk predicted in \cite{Co05} is just a spatially short-scale and very rapid transient phenomenon. This is the base of our criticism to the Shakura prescription, because if the magnetic backreaction is damped by the dissipation required to deal with a satisfactory accretion, where does the current able to ensure the proper azimuthal GOL come from? 

We conclude observing the the solution to this puzzling picture can not come from the interaction of the disk with its environment, for instance with the disk corona \cite{UG08, KB04}. In fact, the process responsible for exchange of energy between the disk and its corona seems to be driven, on a magnetic level, on very large spatial scales \cite{BP09}. The backreaction we studied here, in the extreme cases, is a phenomenon having the spatial scale of the meters and the typical duration of $\mu$seconds. It is just this peculiar feature which allows us to claim firmly the presence of an inconsistent scheme, as soon as, the hydrodynamical and magnetic scale are separated to recover the Shakura profile. Indeed, it is clear that such a scheme can not be affected by the typical space and time scales of the fundamental physical processes governing the disk profile evolution. Therefore, only two possible scenarios remain viable in the perspective of the present issue: \emph{(i)} it requires an enormous anomalous value of the resistivity coefficient, as discussed above; \emph{(ii)} the separation of the hydrodynamical and the magnetic scales is removed and a deep interchange between the kinetic and magnetic energy of the disk takes place (the disk morphology is then rather far from the Shakura prescription). This second perspective requires the set-up of sophisticated numerical simulations, able to access the very short space and time scale, whose contribution can not be, in principle, disregarded as outlined here.

{\small ** This work was developed within the framework of the \emph{CGW Collaboration} (www.cgwcollaboration.it). NC gratefully acknowledges the CPT - Universit\'e de la Mediterran\'ee Aix-Marseille 2 and the financial support from ``Sapienza'' University of Roma. **}

\end{document}